\documentclass[preprint]{aastex}
\usepackage{amsmath}
\usepackage{cases}
\usepackage{url}
\usepackage{color}
\usepackage{lineno}

\shorttitle{Mutli-Scale Pressure-balanced Structures}
\shortauthors{Yang et al.}

\begin{document}


\title{Multi-scale Pressure-balanced Structures in Three-dimensional Magnetohydrodynamic Turbulence}


\author{Liping Yang\altaffilmark{1,2}, Jiansen He\altaffilmark{2},
Chuanyi Tu\altaffilmark{2}, Shengtai Li\altaffilmark{3}, Lei Zhang\altaffilmark{1}, Eckart
Marsch\altaffilmark{4}, Linghua Wang\altaffilmark{2}, Xin Wang\altaffilmark{5}, and Xueshang Feng\altaffilmark{1}}


\altaffiltext{1}{SIGMA Weather Group, State Key Laboratory for Space
Weather, National Space Science Center, Chinese
Academy of Sciences, 100190, Beijing, China}
\altaffiltext{2}{School
of Earth and Space Sciences, Peking University, 100871 Beijing,
China; E-mail: jshept@gmail.com}
\altaffiltext{3}{Theoretical Division, MS B284, Los Alamos National Laboratory, Los Alamos, NM 87545, USA}
\altaffiltext{4}{Institute for
Experimental and Applied Physics, Christian Albrechts University at
Kiel, 24118 Kiel, Germany}
\altaffiltext{5}{School of Space and Environment, Beihang University, 100191 Beijing, China}



\begin{abstract}
Observations of solar wind turbulence indicate
the existence of multi-scale pressure-balanced structures (PBSs) in the solar wind. In this work, we conduct a numerical simulation to investigate
multi-scale PBSs and in particular their formation in compressive MHD turbulence. By the use of a higher order Godunov code Athena,
a driven compressible turbulence with an imposed uniform guide field is simulated. The simulation results show that both the magnetic pressure and the thermal pressure
exhibit a turbulent spectrum with a Kolmogorov-like power law, and that in many regions of the simulation domain they are anti-correlated.
The computed wavelet cross-coherence spectrum of the magnetic pressure and the thermal pressure, as well as their space series,  indicate the existence of multi-scale PBSs, with
the small PBSs being embedded in the large ones. These multi-scale PBSs are likely to be related with the highly oblique-propagating slow-mode waves, as the traced multi-scale PBS is found to be traveling in a certain direction at a speed consistent with that predicted theoretically for a slow-mode wave propagating in the same direction.

\end{abstract}




%
%

\section{INTRODUCTION}
In the solar wind, pressure-balanced structures (PBSs), which show an inherent anti-correlation between the fluctuations of the magnetic pressure ($P_\textrm{mag}$) and the thermal pressure ($P_\textrm{th}$),
have often been observed in the past decades and
are considered as major ingredients of solar
wind compressible turbulence (see the review by  \citet{Tu1995} and references
therein). For observations near 1 AU, \cite{Burlaga1970}
examined the correlation between magnetic and thermal pressures and found that they are negatively correlated, indicating the
existence of PBSs.
In the Voyager data, \citet{Burlaga1990} identified that the PBSs in the outer heliosphere increase with the heliocentric distance.
Using the Helios data, \cite{Marsch1993} and \cite{Tu1994} extensively studied
the correlations between such quantities as density, temperature, and magnetic field magnitude. They found an anti-correlation between the magnetic pressure and the thermal pressure for fluctuations at scales less than one hour.
On the basis of Ulysses observations of the high-latitude solar wind, \cite{Reisenfeld1999} showed that at time scales of less than 1 day, PBSs dominate the compressive component of the plasma turbulence of the high-latitude solar wind (see also \cite{McComas1996}), and
variations of the plasma $\beta$ within PBSs appear to be positively correlated with the variations of the helium abundance. In high-resolution Cluster data,
\citet{Kellogg2005} found that PBSs are possible on the scale of seconds. \cite{Yao2011} selected a segment of data obtained by the Cluster C1 spacecraft in quiet solar wind, and found a dominant anti-correlation between
the time series of the electron density and magnetic field strength, indicating
the existence of multi-scale PBSs, whose durations can last from 1 hr down to 10 s.

Though PBSs were  frequently identified in the  multi-scale fluctuations in the solar wind,
their nature and origin  have not yet been understood.  In the inner heliosphere, the PBSs
progressively  fade away with increasing heliocentric distance, which suggests that they may be of a solar origin as ``spaghetti-like'' magnetic flux tubes \citep{Mariani1973, Thieme1989, Thieme1990, Borovsky2008}, while in the outer heliosphere
the PBSs build up increasingly, which means that
they may also be generated in the solar wind, such as  by slow-mode magnetosonic waves or mirror-mode instability
\citep{Tu1994, Kellogg2005, Yao2013a, Yao2013b}, or by pick-up ions \citep{Burlaga1994}.  \cite{Thieme1989} used the data measured by the two Helios spacecrafts to investigate properties of high-speed streams,
and presented clear cases that the variability of the fluid parameters, which characterizes the fine-stream tubes, are due to spatial structures that may be viewed as remnants of the underlying coronal structures.  \cite{Thieme1990} further pointed out that these fine-stream structures are likely to be pressure-balanced structures. However, this solar-origin explanation is challenged by the recent observation of PBSs with small scales of seconds, as their source region on the Sun  when being mapped out to the heliosphere would be too small to be
observable. By analyzing the plasma and magnetic field data obtained by WIND, \cite{Yao2013a} investigated  the dependence of the properties of small-scale PBSs on the background magnetic field direction, and showed that the small-scale PBS may be related with  highly oblique-propagating, slow-mode waves in the solar wind.
Based on the proton temperatures T$_\textrm{perp}$ and T$_\textrm{para}$, \cite{Yao2013b} also gave another example where  the small-scale PBS in the quiet solar wind may have been formed due to mirror-mode instability. The complete evidences of oblique/quasi-perpendicular propagating slow magnetosonic waves are provided by \citet{He2015ApJ}, in which not only the anti-correlation between $\left|B\right|$ and density is presented but also the phase relation between density and parallel velocity ($V_\parallel$) is investigated helping to diagnose the propagation direction with respect to the magnetic field vector.

Apart from these observational evidences, numerical simulations have been conducted to investigate the nature of the PBSs and discontinuties \citep{Greco2008, Greco2009, Servidio2011, Zhdankin2012a, Zhdankin2012b, Verscharen2012, Yang2015, Zhang2015b}.  By comparing solar wind data from the Advanced Composition Explorer (ACE)  with simulations of
magnetohydrodynamic (MHD) turbulence, \citet{Greco2009} examined the link between discontinuity identification and intermittency analysis, which supports
that  many or most of the intermittent solar wind structures are discontinuities which are expected to be
produced in MHD turbulence.  \cite{Verscharen2012} presented a two-dimensional model of solar wind turbulence in the dissipation range, and illustrated some signatures of small-scale PBSs that are associated with very oblique slow-mode
waves. In a simulation of the three-dimensional (3-D) decaying  compressive MHD turbulence, \cite{Yang2015} examined how the
rotational discontinuities  are formed. They found that rotational discontinuities can evolve from the steepening of the Alfv\'{e}n wave
with moderate amplitude, and this steepening is caused by the inhomogeneity of the Alfv\'{e}n speed in the ambient turbulence.
However, these works have not reported multi-scale PBSs, especially in the sense that the small-scale PBSs are embedded in the larger ones. Their possible formation in MHD turbulence has not been demonstrated.

Yet in the present work, based on a simulation of the driven compressible MHD turbulence, we identified frequent occurrences of multi-scale PBSs, and could attribute their formation to oblique-propagating, slow-mode
waves. In Section 2, a general description of the numerical MHD
model is given. Section 3 describes the results of the numerical
simulation and their analysis. Section 4 is reserved for the summary
and discussion.

\section{NUMERICAL MHD MODEL}

\begin{figure}[htbp]
   \begin{center}
   \begin{tabular}{c}
     \includegraphics[width=6.5  in]{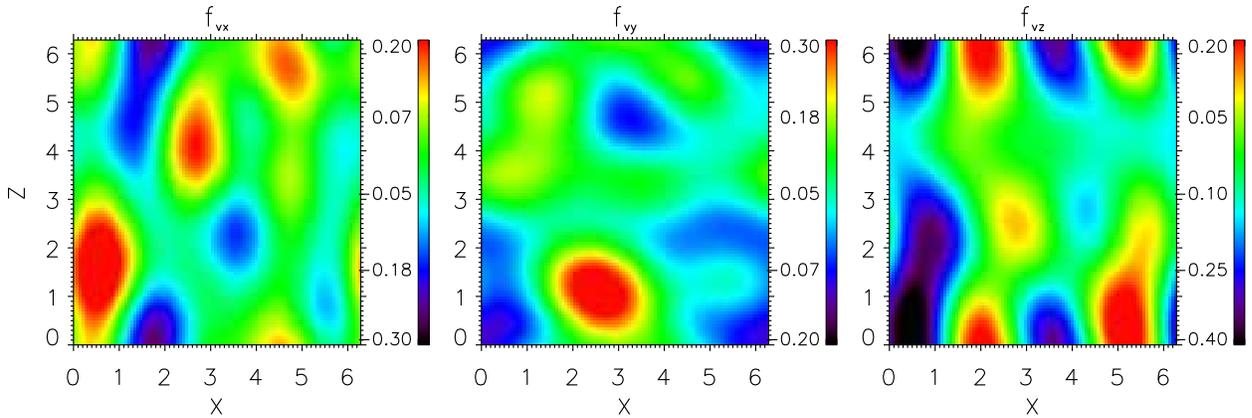}   \\
  \end{tabular}
   \end{center}
\caption{Distributions of the random large-scale drivers $f_{vx}$, $f_{vy}$, and $f_{vz}$  in the $y=3.14$ plane at t = 0. }\label{figure1}
\end{figure}

\begin{figure}[htbp]
   \begin{center}
   \begin{tabular}{c}
     \includegraphics[width=2.5  in]{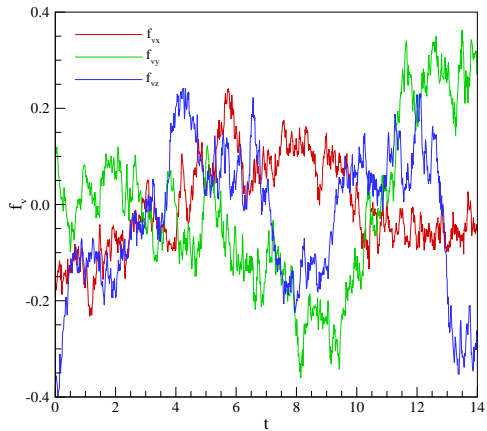}   \\
  \end{tabular}
   \end{center}
\caption{Temporal evolution of the random large-scale drivers $f_{vx}$, $f_{vy}$, and $f_{vz}$  at $x=0.63$, $y=0.63$, and $z=0.63$. }\label{figure2}
\end{figure}

The plasma is described by a compressible 3-D MHD model,
which involves fluctuating variables and a uniform large-scale field $B_0$  in the $z-$direction. The equations are written in the following non-dimensional form:

\begin{equation}
 \frac{\partial \rho}{\partial
t}+ \nabla \cdot (\rho \mathbf{u}) = 0 \ ,
\end{equation}
\begin{equation}
 \frac{\partial \rho \mathbf{u}}{\partial
t}+ \nabla \cdot \left[\rho \mathbf{u} \mathbf{u} + ( p +
\frac{1}{2}\mathbf{B}^2 )\mathbf{I}-\mathbf{B} \mathbf{B}\right] = \mathbf{f}_v \ ,
\end{equation}
\begin{equation}
 \frac{\partial e}{\partial
t}+ \nabla \cdot \left[\mathbf{u} (e + p +
\frac{1}{2}\mathbf{B}^2)-(\mathbf{u} \cdot
\mathbf{B})\mathbf{B}\right]= \nabla \cdot (\mathbf{B} \times \eta \mathbf{j}),
\end{equation}
\begin{equation}
\frac{\partial \mathbf{B}}{\partial t}+ \nabla \cdot
(\mathbf{u}\mathbf{B}-\mathbf{B}\mathbf{u}) = \eta\nabla^2
\mathbf{B} +  \mathbf{f}_b \ ,
\end{equation}
where
\begin{equation}
e=\frac 12 \rho\mathbf{u}^2+\frac{p}{\gamma-1}+\frac 12
\mathbf{B}^2, \ \ \ \ \mathbf{j}  = \nabla \times \mathbf{B}
,\end{equation} which correspond to the total energy density and
current density, respectively. Here, $\rho$ is the mass density;
$\mathbf{u}=(v_x, v_y, v_z)$ is the flow velocity composed of $x$, $y$, and $z-$components; $p$ is the thermal pressure; $\mathbf{B}$ denotes the
magnetic field; $t$ is time; $\gamma= 5/3$ is the adiabatic index;
$\eta$ is the magnetic resistivity; and $\mathbf{f}_v$ as well as  $\mathbf{f}_b$ are the random large-scale drivers, and the distributions of $\mathbf{f}_v$  in the $y=3.14$ plane at t = 0 is shown in Figure \ref{figure1}. An example of the temporal behavior of $\mathbf{f}_v$ at
$x=0.63$, $y=0.63$, and $z=0.63$ is presented in Figure \ref{figure2}, and the amplitude of $\mathbf{f}_{vx}$, $\mathbf{f}_{vy}$, and $\mathbf{f}_{vz}$ at the high frequency is about 0.05, which is far smaller than the magnetic(thermal) pressure gradient at the grid size (i.e. $\nabla P_\textrm{mag(th)}$) as shown in Figure \ref{figure4}. Therefore, the behaviors of $\mathbf{f}_v$ and $\mathbf{f}_b$ do not affect the generation of PBSs directly without nonlinear cascading process.

\begin{figure}[htbp]
   \begin{center}
   \begin{tabular}{c}
     \includegraphics[width=3 in]{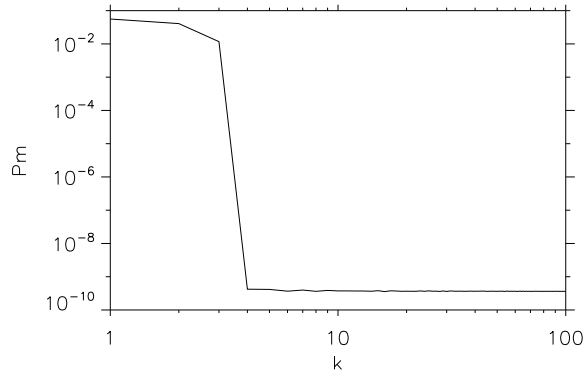}   \\
  \end{tabular}
   \end{center}
\caption{Profile of $Pm$ with $k$. }\label{figure3}
\end{figure}

Specifically, the components of $\mathbf{f}_v$ and $\mathbf{f}_b$ are defined in Fourier space as $Pm(k)\exp(i\phi)$, where the profile of $Pm(k)$ is shown in Figure \ref{figure3}, and $\phi$ is the uniform-distributed random phase angle between [0, 2$\pi$]. Both $\mathbf{f}_v$ and $\mathbf{f}_b$ are non-compressive (although the driven turbulence is compressive), satisfying $\nabla \cdot \mathbf{f}_v = 0$ and $\nabla \cdot \mathbf{f}_b = 0$. To introduce the cross-helicity (indicator of imbalance) and the Alfv\'en ratio less than 1, the amplitude of $\mathbf{f}_b$ is designed to be about 3 times larger than that of $\mathbf{f}_v$.   The driven turbulence is not pure Alfv\'enic, as the direction of forces is demanded to be only perpendicular to the
wave vector $\mathbf{k}$, not the large-scale field $B_0$.

Under typical solar wind conditions, the Alfv\'en ratio is less than 1, and the perturbed magnetic field strength is less than the mean large-scale magnetic field strength. To be comparable to these observations, the amplitudes
of the forcing components are tuned to maintain that the root-mean-square (rms) values for the velocity and the magnetic field
in statistically stationary state is approximately 0.67 and 0.75, respectively. The large-scale field $B_0$ is set to be 1.00 \^z.

We consider periodic boundary conditions in a cube with a side
length of $2\pi$ and a resolution defined by $512^3$ grid points, and run a
simulation from an initial state with only the guide field $B_0$. The initial density and thermal pressure are set to be uniform.
Hence, the simulation domain is initially without fluctuations, and that the turbulence is generated by the drivers alone.

For the calculation presented herein, we employ a higher order Godunov code,  Athena \citep{Gardiner2005, Gardiner2008, Stone2008}. Specifically,
we apply  a third-order piecewise parabolic method to the reconstruction and the
approximate Riemann solver of Harten-Lax-van Leer  Discontinuities (HLLD)
to the calculation of the
numerical fluxes. The constrained transport algorithm is applied for ensuring the divergence-free state of the magnetic field.

\section{NUMERICAL RESULTS}

\begin{figure}[htbp]
   \begin{center}
   \begin{tabular}{c}
     \includegraphics[width=6.5  in]{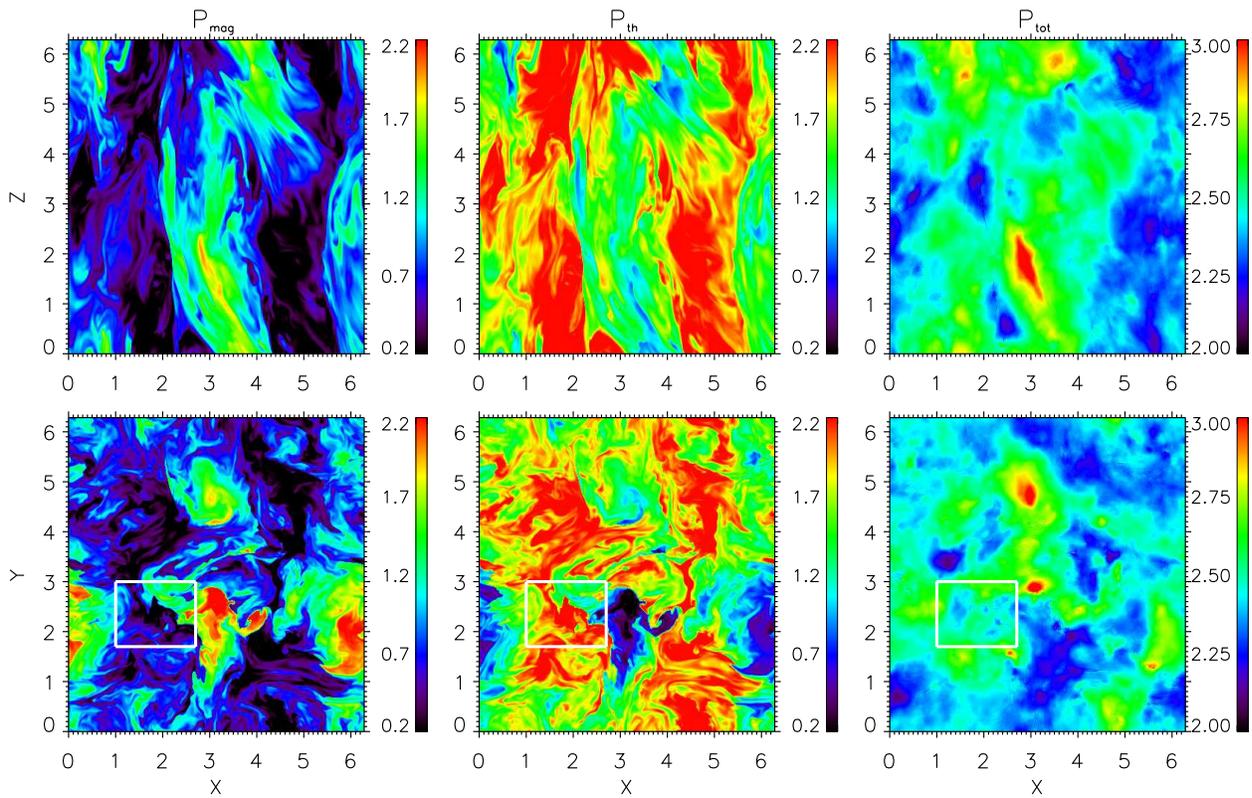}   \\
  \end{tabular}
   \end{center}
\caption{Distributions of the magnetic pressure $P_\textrm{mag}$, the thermal pressure $P_\textrm{th}$, and the total pressure $P_\textrm{tot}$,  in the $y=4.64$ (top) and $z=1.55$ (bottom) planes at
$t=22$. }\label{figure4}
\end{figure}

Figure \ref{figure4} presents the calculated distributions of the magnetic pressure $P_\textrm{mag}$, the thermal pressure $P_\textrm{th}$, and the total pressure $P_\textrm{tot}$,  in the $y=3.14$ and $z=3.14$ planes at
$t=22$. This figure shows that as a result of the well-known anisotropic behavior of fluctuations in MHD with a uniform guide
field, the structures preferentially align along the $z-$direction, and become much more varying in the perpendicular
cross section. Both the magnetic pressure $P_\textrm{mag}$ and the thermal pressure $P_\textrm{th}$  are fluctuating, while the total pressure $P_\textrm{tot}$  changes not so chaotic. Taking the region marked by the white rectangle in this figure as an example, both $P_\textrm{mag}$ and $P_\textrm{th}$ alter from 0.2 to 2.2, and show fine structures, while $P_\textrm{tot}$ is between 2.25 and 2.75, without such fine structures.
In many regions we can see such great changes of both $P_\textrm{mag}$ and $P_\textrm{th}$, yet $P_\textrm{tot}$ only changes sightly.

\begin{figure}[htbp]
   \begin{center}
   \begin{tabular}{c}
     \includegraphics[width = 4 in]{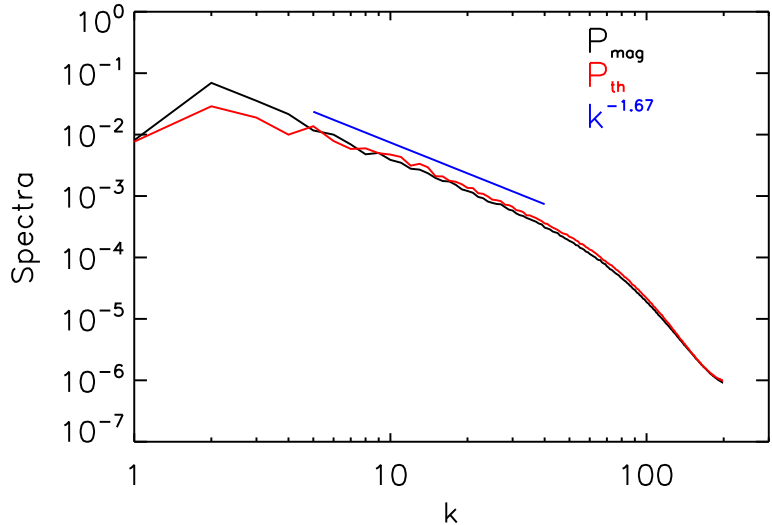}   \\
  \end{tabular}
   \end{center}
\caption{Spectra of the magnetic pressure $P_\textrm{mag}$ (black) and the thermal pressure $P_\textrm{th}$  (red) with a Kolmogorov-like power law spectrum (blue) for reference.}\label{figure5}
\end{figure}

Figure \ref{figure5} shows the spectra of the magnetic pressure $P_\textrm{mag}$  and the thermal pressure $P_\textrm{th}$. To calculate the spectra of  $P_\textrm{mag}$  and $P_\textrm{th}$, we first transform them into 3D Fourier ($\mathbf{k}$) space, and then added together the squared Fourier amplitudes over $\Delta k$ (here set to be 1) within a spherical shell $(k-\Delta k/2, k+\Delta k/2)$, that is, $\frac{\int_{-\Delta k/2}^{\Delta k/2} \int_0^{2\pi} \int_0^\pi \mid FFT(P_{mag(th)})\mid^2 k^2 \sin\theta  \delta \theta \delta \phi \delta k}{
\Delta k}$
Although the simulation models a compressible magnetofluid,
both the magnetic pressure  and the thermal pressure exhibit a turbulent spectrum with a Kolmogorov-like power law that is a characteristic of an incompressible hydrodynamic fluid.
This result is comparable with in situ solar wind observations \citep{Tu1994, Yao2011} and previous simulation \citep{Shaikh2010}. Also, in the inertial range, the spectra of $P_\textrm{mag}$ and $P_\textrm{th}$  are close, which indicates their amplitude change similarly over the various spatial scales.

\begin{figure}[htbp]
   \begin{center}
   \begin{tabular}{c}
     \includegraphics[width=6. in]{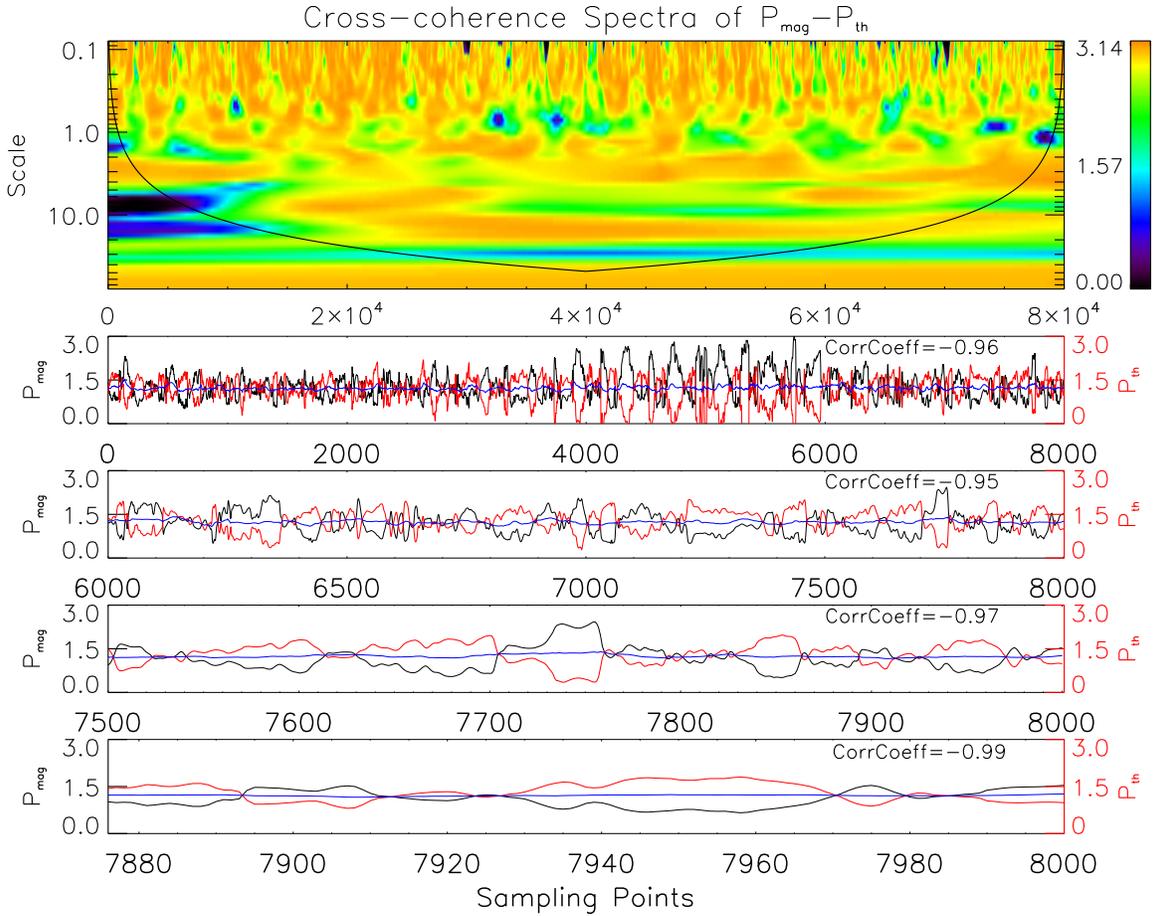}   \\
  \end{tabular}
   \end{center}
\caption{Anti-correlation between the magnetic pressure $P_\textrm{mag}$ and the thermal pressure $P_\textrm{th}$  at different scales. The first panel is the wavelet cross-coherence spectrum
 of $P_\textrm{mag}$ and $P_\textrm{th}$, showing the product of  the coefficient spectrum  and the  absolute phase spectrum. The other panels plot the space series of $P_\textrm{mag}$ (black), $P_\textrm{th}$  (red), and $P_\textrm{tot}/2$  (blue), annotated with correlation coefficients.}\label{figure6}
\end{figure}

To study the possible correlation between $P_\textrm{mag}$ and $P_\textrm{th}$  at different scales, we first sample $P_\textrm{mag}$ and $P_\textrm{th}$  along a straight line,
and then calculate the wavelet cross-coherence spectrum
of $P_\textrm{mag}$ and $P_\textrm{th}$  \citep{Yao2011}. In that research, the data sampled by Cluster C1 spacecraft can be interpreted, on considerations of Taylor's hypothesis, as a sampled result along a line with a mean angle of $45^\circ$ to the interplanetary field. For numerical simulation data reported here, similarly we sample simulated data with a line orienting $45^\circ$ to the direction of the large-scale $B_0$ (\^{z}). We have tried other different orientations, and still detected the multi-scale PBSs.

To study the correlation between the magnetic pressure $P_\textrm{mag}$ and the thermal pressure $P_\textrm{th}$  at different scales, we first calculate the wavelet coefficients $\omega_\textrm{mag}(i, \tau)$ and $\omega_\textrm{th}(i, \tau)$ as a function
of the point $i$ and the scale $\tau$ for $P_\textrm{mag}$ and $P_\textrm{th}$, and then get the coherence coefficient as $ \frac{\mid \omega_\textrm{mag}(i, \tau) \times \omega_\textrm{th}^\ast (i, \tau)\mid ^2}{\mid \omega_\textrm{mag}(i, \tau)\mid^2 \times \mid \omega_\textrm{th}(i, \tau)\mid^2 } $  as well as the  absolute phase as $\textrm{tan}^{-1} (\frac{\textrm{imaginary}(\omega_\textrm{mag}(i, \tau) \times \omega_\textrm{th}^\ast (i, \tau))}{\textrm{real}(\omega_\textrm{mag}(i, \tau) \times \omega_\textrm{th}^\ast (i, \tau))})$ \citep{Torrence1998}.

Figure \ref{figure6} shows the product of  the coherence coefficient spectrum  and the  absolute phase spectrum as well as the space series of $P_\textrm{mag}$, $P_\textrm{th}$, and $P_\textrm{tot}/2$.
This figure highlights the regions with significant anti-correlation, where the coherence coefficient is close to 1 and the absolute phase angle is close to $\pi$. Like the observations in the solar wind \citep{Yao2011}, this anti-correlation extends over the
whole series of points located inside the region framed by the black line. As the scale decreases, the regions with sharp change of $P_\textrm{mag}$ and $P_\textrm{th}$ become smaller, and more PBSs appear. To be mentioned, the black line shown here is defined as the $e-$folding time ($\sqrt{2} \tau$ for Morlet, see Table 1 in \citet{Torrence1998}) for the autocorrelation of wavelet power at
each scale, and confines a bell-shaped area, beyond which the effect of the finiteness of data may influence the wavelet spectra \citep{Torrence1998}.

Along the path we sample 8000 points, and form 4 segments, with smaller ones embedded in larger ones, respectively, to display the variation of $P_\textrm{mag}$, $P_\textrm{th}$, and $P_\textrm{tot}/2$ on different scales.
Like for the solar wind observations \citep{Yao2011}, simulation results reveal that a multi-scale anti-correlation between $P_\textrm{mag}$ and $P_\textrm{th}$  exists, and that multi-scale PBSs are generated, whereby the small PBSs are found to be embedded in the large ones.
On the scale of 8000 points, a strong anti-correlation between $P_\textrm{mag}$ and $P_\textrm{th}$  appears, with a
correlation coefficient of -0.96. When zooming into the last 2000 points, the variations of $P_\textrm{mag}$ and $P_\textrm{th}$  are seen to be also negative, with a correlation coefficient
of -0.95. For the sequent scale of only 500 points, the anti-correlation with the coefficient of -0.97 continues.
Finally, we zoom into an even smaller scale of 125 points, where we find an anti-correlation
with a coefficient of -0.99.

\begin{figure}[htbp]
   \begin{center}
    \begin{tabular}{c}
     \includegraphics[width=5. in]{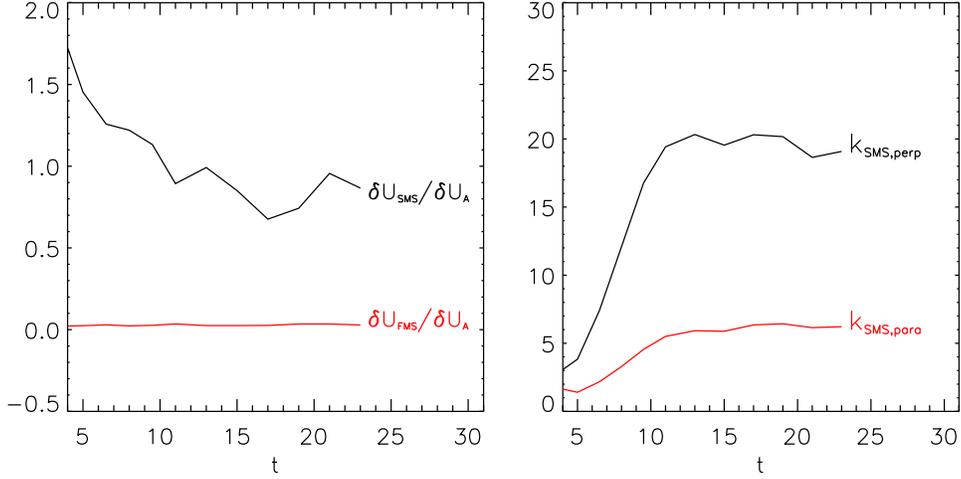}   \\
  \end{tabular}
   \end{center}
\caption{The time evolution of the root-mean-square (rms) perturbation speeds of fast ($\delta U_{\scriptsize\textrm{FMS}}$) as well as slow ($\delta U_{\scriptsize\textrm{SMS}}$) magnetosonic waves relative to that of Alfv\'{e}n waves ($\delta U_{\scriptsize\textrm{A}}$) (Left panel) and the rms
perpendicular wavenumber $K_{\scriptsize\textrm{SMS, perp}}$ as well as parallel wavenumber $K_{\scriptsize\textrm{SMS, para}}$ of the slow magnetosonic waves (Right panel).  }\label{figure7}
\end{figure}

To investigate the nature of waves occurring possibly in the turbulence for each $\mathbf{k}$, the perturbation speeds, corresponding to the fast ($\delta u_{\scriptsize\textrm{FMS}}$) and the slow ($\delta u_{\scriptsize\textrm{SMS}}$) magnetosonic waves and the Alfv\'{e}n waves ($\delta u_{\scriptsize\textrm{A}}$), are computed  via projections, based on the fact that the three MHD wave mode velocities $\delta \mathbf{v_{\scriptsize\textrm{FMS}}} (= v_{\textrm{p, F}}^2 \mathbf{k} - v_{\textrm{A}}^2 \cos \theta  \mathbf{b_0}) $, $\delta \mathbf{v_{\scriptsize\textrm{SMS}}} (= v_{\textrm{p, S}}^2 \mathbf{k} - v_{\textrm{A}}^2 \cos \theta  \mathbf{b_0})$, and $\delta \mathbf{v_{\scriptsize\textrm{A}}} (=\mathbf{k} \times \mathbf{b_0})$ form an orthogonal coordinate system for a given $\mathbf{k}$, where $v_{\textrm{p, F}}$, $v_{\textrm{p, S}}$, as well as $v_{\textrm{A}}$ are the phase speed of the fast and slow magnetosonic waves as well as Alfv\'{e}n waves, and $\theta$ is the angle between unit vector $\mathbf{k}$ and $\mathbf{b_0}$ \citep{Marsch1986, Zhang2015}. Specifically, $ \delta u_{\scriptsize\textrm{FMS}} \sim \mathbf{u(\mathbf{k})} \cdot \delta \mathbf{v_{\scriptsize\textrm{FMS}}}/\mid \delta \mathbf{v_{\scriptsize\textrm{FMS}}}\mid $, $\delta u_{\scriptsize\textrm{SMS}} \sim  \mathbf{u(\mathbf{k})} \cdot \delta \mathbf{v_{\scriptsize\textrm{SMS}}}/\mid \delta \mathbf{v_{\scriptsize\textrm{SMS}}}\mid $, and $\delta u_{\scriptsize\textrm{A}} \sim \mathbf{u(\mathbf{k})} \cdot \delta \mathbf{v_{\scriptsize\textrm{A}}}/\mid \delta \mathbf{v_{\scriptsize\textrm{A}}}\mid  $.
Please note that $\mathbf{u}(\mathbf{k})$ is assumed to be the superposition of velocity fluctuations of three kinds of MHD waves. In fact, a certain fraction of $\mathbf{u}(\mathbf{k})$ may not be associated with any of the three types of MHD waves. Moreover, from the calculated $\delta u_{\scriptsize\textrm{SMS}}$, we define the
rms perpendicular and parallel wavenumber by $K_{\scriptsize\textrm{SMS, perp}} = \sqrt{\sum_k (k_x^2 + k_y^2)  \delta u_{\scriptsize\textrm{SMS}}  / \sum_k \delta u_{\scriptsize\textrm{SMS}}}$, and $K_{\scriptsize\textrm{SMS, para}} = \sqrt{\sum_k (k_z^2) \delta u_{\scriptsize\textrm{SMS}}/ \sum_k  \delta u_{\scriptsize\textrm{SMS}}}$.

In Figure \ref{figure7}, we plot the evolution of the  rms perturbation speeds of fast ($\delta U_{\textrm{\scriptsize{FMS}}}$) as well as slow ($\delta U_{\scriptsize\textrm{SMS}}$) magnetosonic waves relative to that of Alfv\'{e}n waves ($\delta U_{\scriptsize\textrm{A}}$) (Left panel) and the rms
perpendicular wavenumber $K_{\scriptsize\textrm{SMS, perp}}$ as well as parallel wavenumber $K_{\scriptsize\textrm{SMS, para}}$ of the slow magnetosonic waves (Right panel). Figure \ref{figure7} shows that $\delta U_{\scriptsize\textrm{FMS}}$ is marginal compared with $\delta U_{\scriptsize\textrm{SMS}}$ and $\delta U_{\scriptsize\textrm{A}}$, and after the simulated turbulence reaches a statistically stationary state, $\delta U_{\scriptsize\textrm{SMS}}$ approaches $\delta U_{\scriptsize\textrm{A}}$. Therefore, the computation domain is full of slow magnetosonic waves as well as Alfv\'{e}n waves. The compressive fluctuations behave as slow mode waves as the negative correlation between the magnetic pressure and the thermal pressure indicates. The evolution of the $K_{\scriptsize\textrm{SMS, perp}}$ and $K_{\scriptsize\textrm{SMS, para}}$ shows that before $t = 10$, most of the slow magnetosonic waves propagate more and more obliquely, and after that most of them keep propagating highly obliquely.

\begin{figure}[htbp]
   \begin{center}
    \begin{tabular}{c}
     \includegraphics[width=6 in]{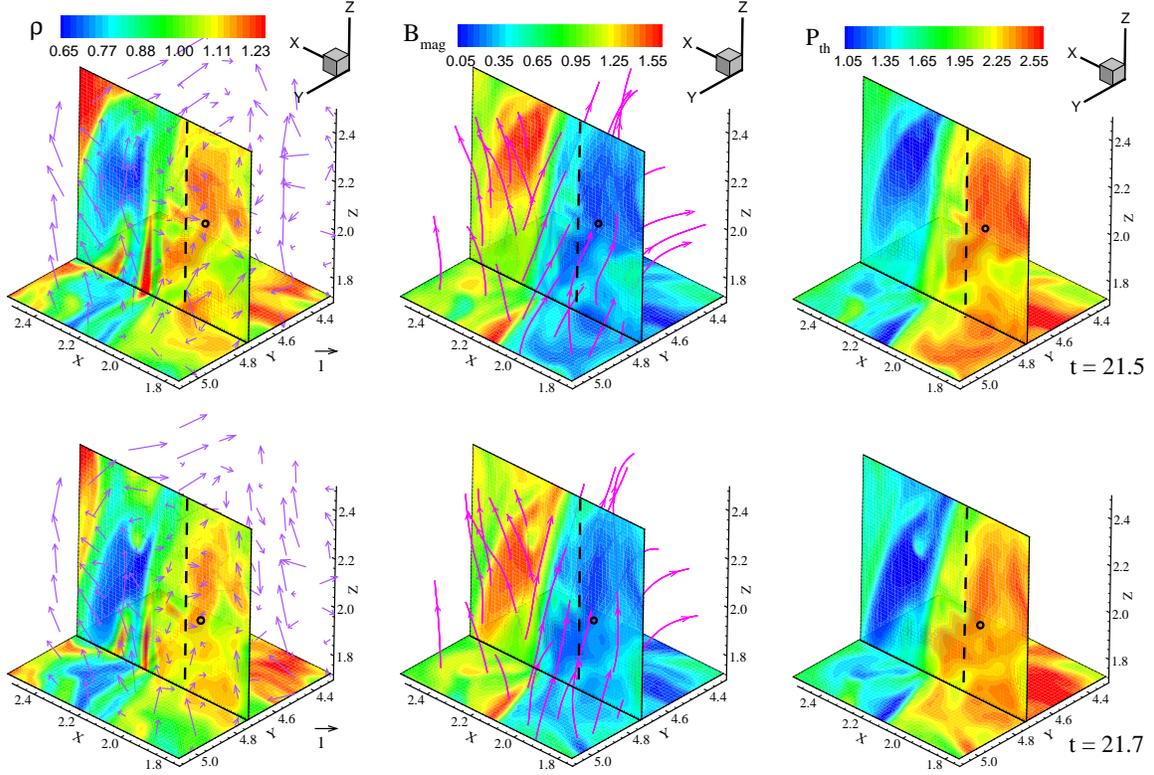}   \\
  \end{tabular}
   \end{center}
\caption{Distributions of the density $\rho$, the magnetic pressure $P_\textrm{mag}$, and the thermal pressure $P_\textrm{th}$  at
$t=21.5$ (the first row) and $t=21.7$ (the second row), respectively. Superposed are by purple arrows the velocity fields, and superposed by pink lines the magnetic field lines. Black cycle marks the movement of the chunk of the high density at the right, and black dashed line is the slice for the time-distance diagrams of Figure \ref{figure9} in the $z-$direction. }\label{figure8}
\end{figure}

To see whether the multi-scale PBSs detected here could be related with the obliquely propagating slow-mode waves,  Figure \ref{figure7} displays the calculated distributions of the density $\rho$, the magnetic pressure $P_\textrm{mag}$, and the thermal pressure $P_\textrm{th}$  at $t=21.5$ (the first row) and $t=21.7$ (the second row), respectively. Superposed are by purple arrows the velocity fields, and superposed by pink lines the magnetic field lines. Black cycle marks the movement of the chunk of the high density at the right.
The chunk of the high density denotes the region where the density
and the thermal pressure at the right of Figure \ref{figure8} are shown in red, and meanwhile the magnetic pressure is blue. From $t=21.5$ to $t=21.7$, we determine its position simply through visual tracing by eyes,
and the propagation direction of it could be deemed to be mainly along the $z-$direction.

\begin{figure}[htbp]
   \begin{center}
    \begin{tabular}{c}
     \includegraphics[width=6.5 in]{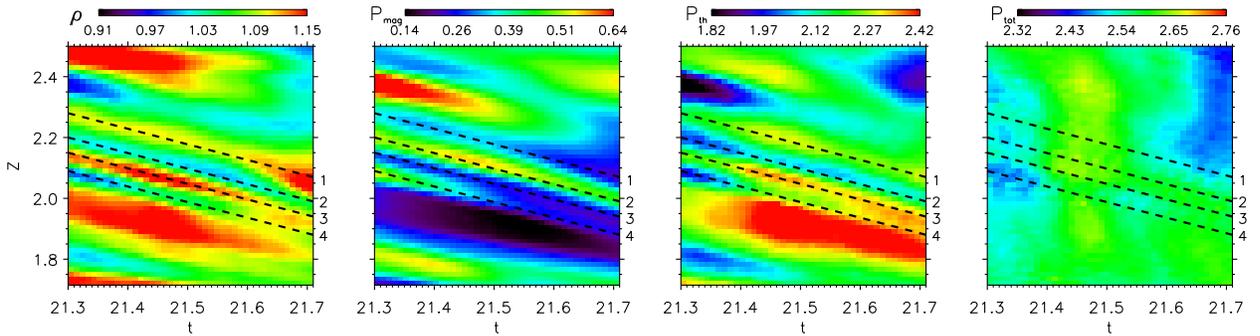}   \\
  \end{tabular}
   \end{center}
\caption{Time-distance diagrams of the density $\rho$, the magnetic pressure $P_\textrm{mag}$, the thermal pressure $P_\textrm{th}$, and the total pressure $P_\textrm{tot}$.
The spatial direction is along the propagation direction defined by the black dashed line in Figure \ref{figure8}, and the slopes of the black dashed lines indicate the propagation
speeds.}\label{figure9}
\end{figure}
To estimate the propagation speed of the chunk of the high density, the values of $\rho$, $P_\textrm{mag}$, $P_\textrm{th}$, and $P_\textrm{tot}$ along the black dashed line in Figure \ref{figure8} are extracted, and are stacked in time sequence. This gives the time-distance
diagrams as displayed in Figure \ref{figure9}, showing the recurrent
stripes, whose slopes correspond to the propagating speeds, and are measure to be about $0.52$. By averaging magnetic fields in the region associated with the chunk of the high density, it could be calculated that the angle between the mean magnetic field and the $z-$direction is  $\theta \sim 50^\circ$. With the use of the average values of thermal pressure and density, we could compute the local slow-mode wave speed along the propagation direction ( i.e. basically the z-direction)  as $\sqrt{\frac{1}{2}[(C_s^2+V_A^2)-\sqrt{(C_s^2+V_A^2)^2-4C_s^2V_A^2\cos^2(\theta)}]}$ ($C_s$ is sonic speed, and $V_A$ is Alfv\'{e}n speed), which is $\sim 0.55$.

Also, at both $t=21.5$  and $t=21.7$, in this region, the velocity is small (less than 0.5), and its directions are diverse, giving a global velocity (averaged over this chunk of the high density at either time point) $\sim 0$. Considering that the magnetic pressure keeps being negatively correlated with the density as well as with the thermal pressure, it can be said that this chunk of the high density is probably associated with  an obliquely propagating, slow-mode wave. At the same time, the magnetic pressure is well in balance with the thermal pressure for this wave, and the multi-scale PBS forms here.

\section{SUMMARY AND DISCUSSION}

In this work, we used a higher order Godunov code to simulate the generation of compressible turbulence in a plasma with an imposed uniform guide field, and to explore the formation of the multi-scale PBSs in compressive MHD turbulence.

The numerical simulation shows that in many regions the magnetic pressure is negatively correlated with the thermal pressure, and both $P_\textrm{mag}$ and $P_\textrm{th}$ alter greatly, yet the total pressure $P_\textrm{tot}$ only changes sightly. $P_\textrm{mag}$ and $P_\textrm{th}$ exhibit a turbulent spectrum with a Kolmogorov-like power law, although this spectrum applies here to compressible turbulence.
In the inertial range, the spectra of the magnetic pressure  and the thermal pressure are close, and the amplitudes of $P_\textrm{mag}$ and $P_\textrm{th}$  at various $k$ are nearly equal.

From the computed wavelet cross-coherence spectrum
of the magnetic pressure and the thermal pressure, it is found that the anti-correlation between them extends over the
whole range of covered scales. As the scale decreases, the regions with sharp changes of $P_\textrm{mag}$ and $P_\textrm{th}$ become small, and more PBSs are generated.  The simulated spatial series of the magnetic pressure  and the thermal pressure
on different scales also show that here exists a multi-scale anti-correlation of them, and that thus multi-scale PBSs are generated, whereby the small PBSs are embedded in the large ones.

By analysis of the nature of waves in the turbulence, it is found that the rms perturbation of the velocity corresponding to the fast magnetosonic waves is marginal compared with that of the slow magnetosonic waves as well as Alfv\'{e}n waves, and most of the compressive fluctuations behave as the slow-mode waves. Also, most of these slow magnetosonic waves reveal a highly oblique propagation.  On the basis of the negative correlation between the magnetic pressure and the density, the negative correlation between the magnetic pressure and the thermal pressure, as well as the approximate equality between the inferred propagation speed and the predicted phase speed of the slow-mode wave, we present one single example that the multi-scale PBS is likely to be related with the highly oblique-propagating slow-mode waves.

In this study, we presented simulation data indicating that the multi-scale PBSs could frequently appear in the compressible turbulence, and that their formation
is likely associated with the oblique-propagating, slow-mode waves. In the simulated turbulence, the slow magnetosonic wave amplitudes are seen to be much stronger than the fast magnetosonic wave amplitudes. This is comparable to the solar wind observation \citep{Tu1994} and earlier simulation \citep{Cho2002}. However, the slow magnetosonic wave amplitudes seem to be comparable to the Alfv\'en wave amplitudes, which is unlike the typical solar-wind conditions that the amplitude in the compressive fluctuations is smaller than the amplitude in the non-compressive Alfv\'enic fluctuations.
This may be related to the nature of the non-pure Alfv\'enic drivers $\mathbf{f}_v$ and $\mathbf{f}_b$. In future, we may investigate the factors that control the generation and evolution of waves as well as their influences on the formation of the multi-scale PBSs.

Another issue concerns the question why the slow-mode waves are highly oblique-propagating. As mentioned by \cite{Chen2012}, is it a result of the strong damping of the parallel-propagating slow-mode wave, or is it determined by the turbulent circumstances? The potential nonlinear interaction between Alfv\'en waves and slow magnetosonic waves, which has been inferred to be existent in the compressible solar wind turbulence and may be related with the observed signature of Landau resonance heating \citep{He2015ApJ}, is another interesting topic worthy of further research. We may answer these questions in our subsequent works.



\begin{acknowledgments}

This work is supported by NSFC grants under contracts 41304133, 41231069, 41574168, 41204105,
 41274132, 41474147, 41421003, and the Specialized Research Fund for State Key Laboratories. The
work was carried out at National Supercomputer Center in Tianjin, China and the calculations were
performed on TianHe-1 (A).

\end{acknowledgments}

\bibliographystyle{apj}

\end{document}